\documentclass[runningheads]{cl2emult}

\usepackage{makeidx}  
\usepackage{graphicx} 
\usepackage{subeqnar} 
\usepackage{multicol} 
\usepackage{cropmark} 
\usepackage{math}     
\makeindex            

\begin{document}
\title*{\bf Dynamics of correlations in the stock market}
\toctitle{Dynamics of correlations in the stock market}
%
%
\titlerunning{Dynamics of correlations in the stock market}
%
\author{S.~Dro\.zd\.z\inst{1,2}
\and F.~Gr\"ummer\inst{1}
\and F.~Ruf\inst{3}
\and J.~Speth\inst{1}}
\authorrunning{S.~Dro\.zd\.z et al.}
%
%
\institute{Institut f\"ur Kernphysik, Forschungszentrum J\"ulich,
D-52425 J\"ulich, Germany
\and Institute of Nuclear Physics, PL-31-342 Krak\'ow, Poland
\and WestLB International S.A., 32-34 bd Grande-Duch. Charl.,\\
L-2014 Luxembourg}

\maketitle              

\begin{abstract}
\index{abstract} Financial empirical correlation matrices of all the
companies which both, the Deutsche Aktienindex (DAX) and the Dow
Jones comprised during the time period 1990-1999 are studied using a
time window of a limited, either 30 or 60, number of trading days.
This allows a clear identification of the resulting correlations. On
both these markets the decreases turn out to be always accompanied by
a sizable separation of one strong collective eigenstate of the
correlation matrix, while increases are more competitive and thus
less collective. Generically, however, the remaining eigenstates of
the correlation matrix are, on average, consistent with predictions
of the random matrix theory. Effects connected with the world
globalization are also discussed and a leading role of the Dow Jones
is quantified. This effect is particularly spectacular during the
last few years, and it turns out to be crucial to properly account
for the time-zone delays in order to identify it. 
\end{abstract}

\section{Introduction} 

Quantifying correlations amoung various financial assets is of great
importance for both practical and fundamental reasons. Practical reasons
point primarily to the theory of optimal portofolio and risk management
(Markowitz 1959, Elton and Gruber 1995).  The fundamental ones, on the
other hand, can also be linked to our understanding of the general
mechanism of time-evolution of complex self-organizing dynamical systems.
One principal characteristics of such a time-evolution is a permanent
coexistence and competition between noise and collectivity. Noise seems to
be dominating, and therefore it is natural that the majority of
eigenvalues of the stock market correlation matrix agree very well (Laloux
et al.~1999, Plerou et al.~1999) with the universal predictions of random
matrix theory (Mehta 1991). Collectivity on the other hand is much more
subtle, but it is this component which is of principal interest, because
it accounts for system-specific non-random properties and thus potentially
encodes the system's future states. In the correlation matrix formalism,
collectivity can be attributed to deviations from the random matrix
predictions. A related recent study (Dro\.zd\.z et al.~2000) demonstrates a
nontrivial time-dependence of the resulting correlations. Generically, the
drawdowns are found always to be accompanied by a sizeable separation of
one strong collective eigenstate of the correlation matrix which, at the
same time, reduces the variance of the noise states. The drawups, on the
other hand, turn out to be more competitive. In this case the dynamics
spreads more uniformly over the eigenstates of the correlation matrix.
Below some of the results of our recent study documenting such effects are
presented.

\section{Financial correlation matrix}

For an asset labelled with $i$ and represented by the price time-series
$x_i(t)$ of length $T$ one defines a time-series of normalized returns
\begin{equation}
g_i(t) = {{G_i(t) - \langle G_i(t) \rangle_t} \over v^2},
\label{eq:g}
\end{equation}
where
\begin{equation}
G_i(t) = \ln x_i(t+\tau) - \ln x_i(t) \approx
{{x_i(t+\tau) - x_i(t)} \over x_i(t)}
\label{eq:Gg}
\end{equation}
are unnormalized returns,
$v = \sigma(G_i)$
is the volatility of $G_i(t)$ and $\tau$ denotes the time lag imposed.
For $N$ stocks the corresponding time series $g_i(t)$ of length $T$
are then used to form an $N \times T$ rectangular matrix $\bf M$.
The correlation matrix is then defined
as
\begin{equation}
{\bf C} = {1\over T} {\bf M} {\tilde {\bf M}}.
\label{eq:C}
\end{equation}
Diagonalizing $\bf C$
$({\bf C} {\bf v}^{\alpha} = {\lambda}_{\alpha} {\bf v}^{\alpha})$
gives the eigenvalues $\lambda_{\alpha}$ $({\alpha}=1,...,N)$
and the corresponding eigenvectors ${\bf v}^{\alpha} = \{v^{\alpha}_i\}$.

A useful null hypothesis corresponds to the case of entirely random
correlations. For the density of eigenvalues $\rho_C (\lambda)$
one then obtains (Sengupta and Mitra 1999) :
\begin{equation}
\rho_C(\lambda) = {T \over {2 \pi \sigma^2 N}}
{\sqrt{(\lambda_{max} - \lambda)(\lambda - \lambda_{min})} \over
{\lambda}},
\label{eq:rhonh}
\end{equation}
and
\begin{equation} 
{\lambda}^{max}_{min} = \sigma^2 (1 + {N/T} \pm 2 {N/T}),
\label{eq:lmn}
\end{equation}
with $\lambda_{min} \le \lambda \le \lambda_{max}$, $N \le T$, and
$\sigma^2$ is equal to the variance of the time series.

Another useful reference  relates to matrices whose
entries are Gaussian distributed, but with the Gaussian centered at a
certain
nonzero value $\gamma \le 1$. Schematically, such a matrix $\bf C$ can
then be
represented as
\begin{equation}
{\bf C} = {\bf G} + \gamma {\bf U},
\label{eq:CGU}
\end{equation}
where $\bf G$ denotes a Gaussian distributed matrix centered at zero
and $\bf U$ denotes a matrix whose entries are all unity.
\begin{figure}
\centering
\includegraphics[width=1.0\textwidth]{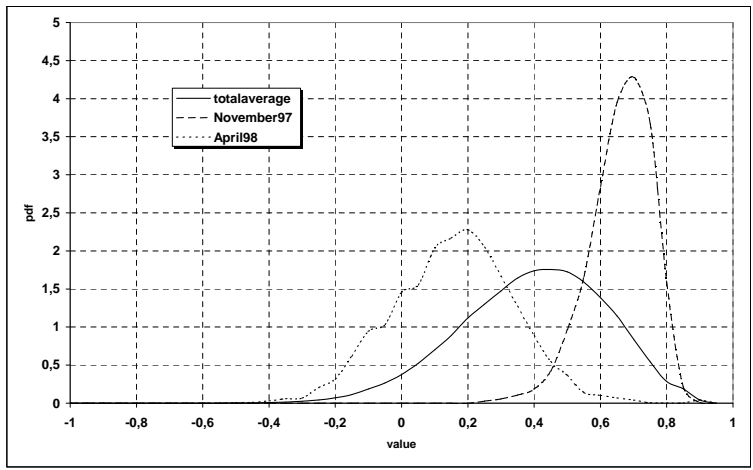}
\caption[]{Distribution of matrix elements of the correlation matrix ${\bf
C}$ (solid line) calculated from the daily price variation of all $N=30$
companies comprised by DAX. The solid line corresponds to an average over
all time windows of length $T=30$ trading days during the period
1990-1999, the dashed line to an average over all such time windows which
end during November 1-30, 1997 and the dotted line to an average over
those which end during April 1-30, 1998.}
\label{fig1}
\end{figure}
\begin{figure}
\centering
\includegraphics[width=1.0\textwidth]{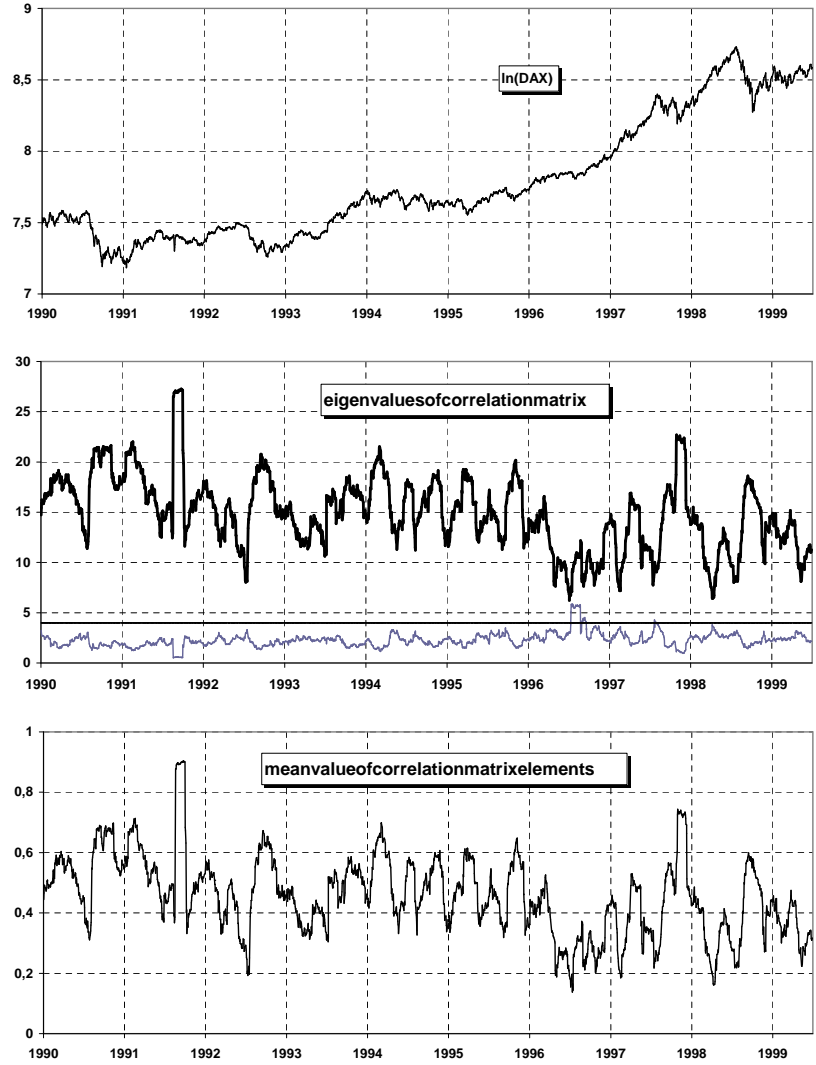}
\caption[]{Upper panel: ln(DAX) during the period 1990-1999. Middle panel:
Time-dependence of two largest eigenvalues corresponding to the DAX
correlation matrix ${\bf C}$ calculated from the time-series of daily
price changes in the interval of $T=30$ past days, during the years
1990-1999. $\lambda_{max}=4$ is indicated by the solid horizontal line.
Lower panel: Time-dependence of the mean value of $\bf C$, which is a
measure for the value of $\gamma$ in eq.~(\ref{eq:CGU}),  during the same
period.}
\label{fig2}
\end{figure}
The rank of $\bf U$ is one and, consequently, the second term alone of the
above equation develops exactly one nonzero eigenvalue of magnitude
$\gamma$. Since the expansion coefficients of this particular state are
all equal, this assigns a maximum of collectivity to such a state. If
$\gamma$ is significantly larger than zero, the dominant structure of $\bf
C$ is expected to be determined by this second term and $\bf G$ can be
considered as just a 'noise' correction. An anticipated result is that one
collective eigenstate with a large eigenvalue is separated by a sizable
gap from the remaining small eigenvalues.

\section{Dynamics of DAX and Dow Jones}

The study presented here is based on daily price variation of all the 
stocks of the Deutsche Aktienindex (DAX) and, independently, of the Dow
Jones during the years 1990-1999. Both these markets involve the same
number $(N=30)$ of the companies and, in this respect, are thus
comparable. The time-dependence of correlations encoded in $\bf C$ can
then be investigated by setting the time window $T=30$ ($T$ should not be
smaller than $N$ as that would artificially reduce the rank of $\bf C$)
and continuously moving it over the whole time period of interest. A clear
indication that the character of correlations may nontrivially vary in
time comes already from the distribution of entries of $\bf C$ recorded at  
different time intervals. Some examples of the corresponding probability  
functionals (pdf) for the DAX are shown in Fig.~\ref{fig1}. In all the
cases the distributions are Gaussian-like, but their variance and location
is significantly different.

The middle panel of Fig.~\ref{fig2} shows the time-dependence of the two
largest eigenvalues of the DAX correlation matrix. Their changes in time
are strong indeed. However, already the second eigenvalue remains
essentially all the time within the bounds prescribed by
eq.~(\ref{eq:lmn}) (in the present case $\lambda^{max}=4$). It is
interesting to relate the largest eigenvalue $(\lambda_1)$ to the global
index (upper panel). As it is quite clearly seen, the global increases and
decreases, respectively, are governed by dynamics of significantly
distinct nature. The decreases are always dominated by one strongly
collective eigenvalue. By conservation of the trace of $\bf C$ the
remaining eigenvalues are then suppressed. The opposite applies to the
increases; their propagation is always less collective. This can be
concluded from the fact that the largest eigenvalue moves down and this
move down is compensated by a simultaneous elevation of the lower
eigenvalues. It is also very interesting to notice that the interpretation
of these results in terms of eq.~(\ref{eq:CGU}) is of an amazing accuracy.
The lower panel of Fig.~\ref{fig2} shows the time-dependence of the mean
value of $\bf C$, the quantity which provides a measure of
$\gamma$. Any differences between the two lines (middle and lower
panel) can hardly be detected.

Another quantity, oriented more towards characterising the localization 
properties of eigenvectors, originates from the information entropy
$I_{\alpha}$ of an eigenvector ${\bf v}^{\alpha}$, defined by its
components as 
\begin{equation}
I_{\alpha} = - \sum_{i=1}^N (v^{\alpha}_i)^2 \ln (v^{\alpha}_i)^2.
\label{eq:ie}
\end{equation}
Using this quantity we then determine a degree of localization
\begin{equation}
L_{\alpha} = \exp (I_{\alpha}).
\label{eq:loc}
\end{equation}
In the case of uniform distribution $(v^{\alpha}_i = 1/\sqrt{N})$ it
yields $L_{\alpha} = N$, i.e., the eigenstate is maximally delocalized.
Another relevant limit is the one which corresponds to a Gaussian
orthogonal ensemble (GOE) of random matrices. In this case
$I_{\alpha}^{GOE} = \psi(N/2 + 1) - \psi(3/2)$ (Izrailev 1990), where
$\psi$ is the digamma function. In our case of $N=30$ this gives $I^{GOE}
\approx 2.67$ and thus $L^{GOE} \approx 14.44$.
\begin{figure}
\centering
\includegraphics[width=1.0\textwidth]{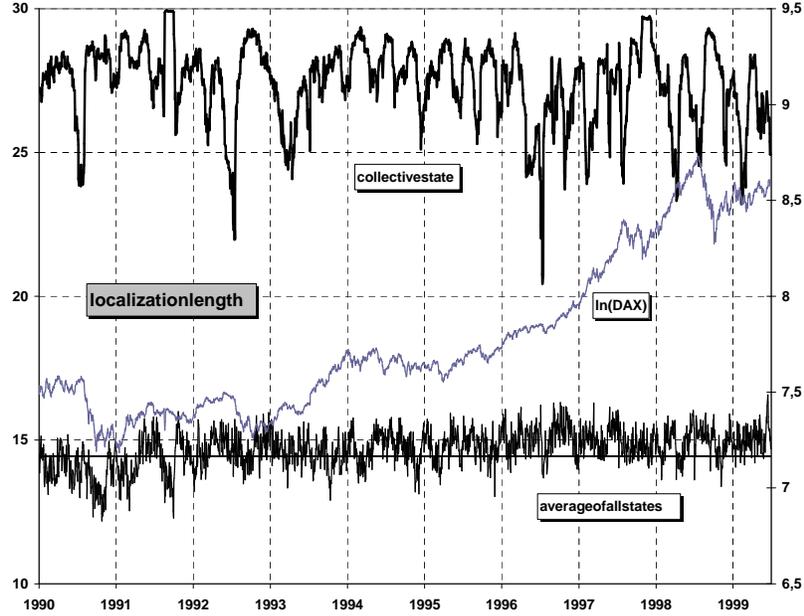}
\caption[]{Time-dependence of the localization length for the most
collective state and for the average over all the states. The solid
horizontal line at $L \approx 14.44$ corresponds to the Gaussian
orthogonal ensemble (GOE) limit and $L=30$ corresponds to a uniformly
distributed eigenvector. In this plot the left vertical axis scale
relates to the localization length, whereas the right axis scale belongs
to the logarithm of the DAX index.}
\label{fig3}
\end{figure}

The two related localization lengths are shown in Fig.~\ref{fig3}.The
upper line corresponds to the collective state and the lower one to the
average taken over all the $I_{\alpha}$'s. As one can see, it happens only
during decreases that the dynamics of the most collective state approaches
the most delocalized form. Otherwise this state becomes more localized. At
the same time the localization length $\langle L \rangle$ corresponding to
$\langle I_{\alpha} \rangle$ oscillates around its GOE limit.  A more
careful inspection shows, however, quite systematic deviations.
Interestingly, on average, $\langle L \rangle$ moves in opposite direction
relative to $I_1$, even though this most collective state is included in
$\langle L \rangle$. This indicates that the stock market drawups lead to
an increase of a global localization length and corrections of the stock
market reduce it. This provides another argument in favour of interpreting
the stock market changes from an increasing to decreasing phase as analogs
of second order phase transitions (Dro\.zd\.z et al.~1999).

\begin{figure}
\centering
\includegraphics[width=1.0\textwidth]{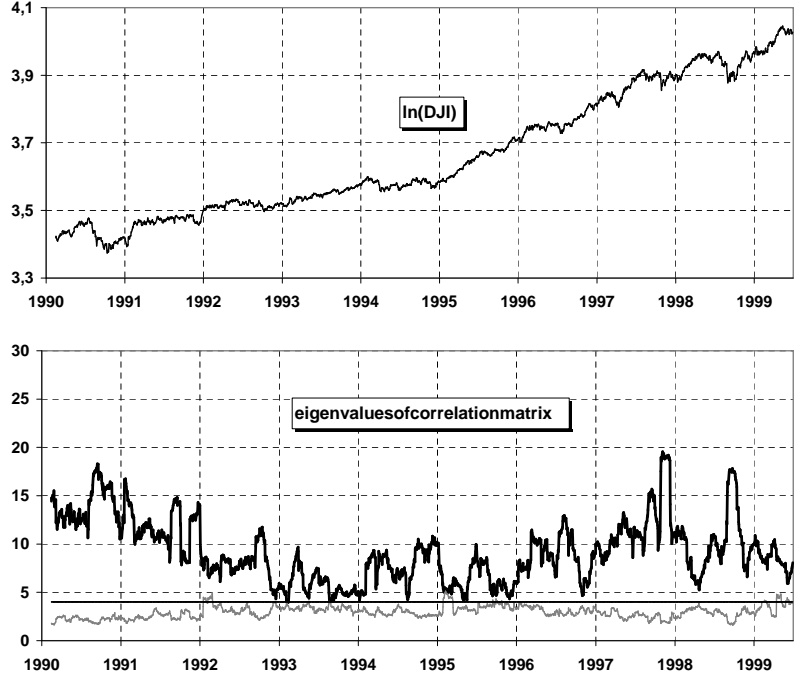}
\caption[]{Same as the two upper panels of Fig.~\ref{fig2} but now for the
Dow Jones.}
\label{fig4}
\end{figure}

Similar conclusions can be drawn form an analogous study of correlations
amoung all the Dow Jones companies. One interesting difference is that on
avarage the dynamics is here less collective. The magnitude of the
separation between the largest eigenvalue and the remaining ones is
systematically smaller for the Dow Jones than for the DAX, as can be
easily seen by comparing Fig.~\ref{fig4} to Fig.~\ref{fig2}.

\section{DAX versus Dow Jones cross-correlations}

All the above mentioned results are based on studies of single stock
markets, in isolation to all others. In view of an increasing role
of effects connected with the world globalization which, as every day
experience indicates, seems to affect also the financial world, it is of
great interest to quantify the related characteristics.
Below we therefore study the cross-correlations
between all the stocks comprised  by DAX and by Dow Jones.
Mixing them up results in 60 companies which determines the size of the
correlation matrix to be studied. Consequently the time window $T=60$ is
also used.
\begin{figure}
\centering
\includegraphics[width=1.0\textwidth]{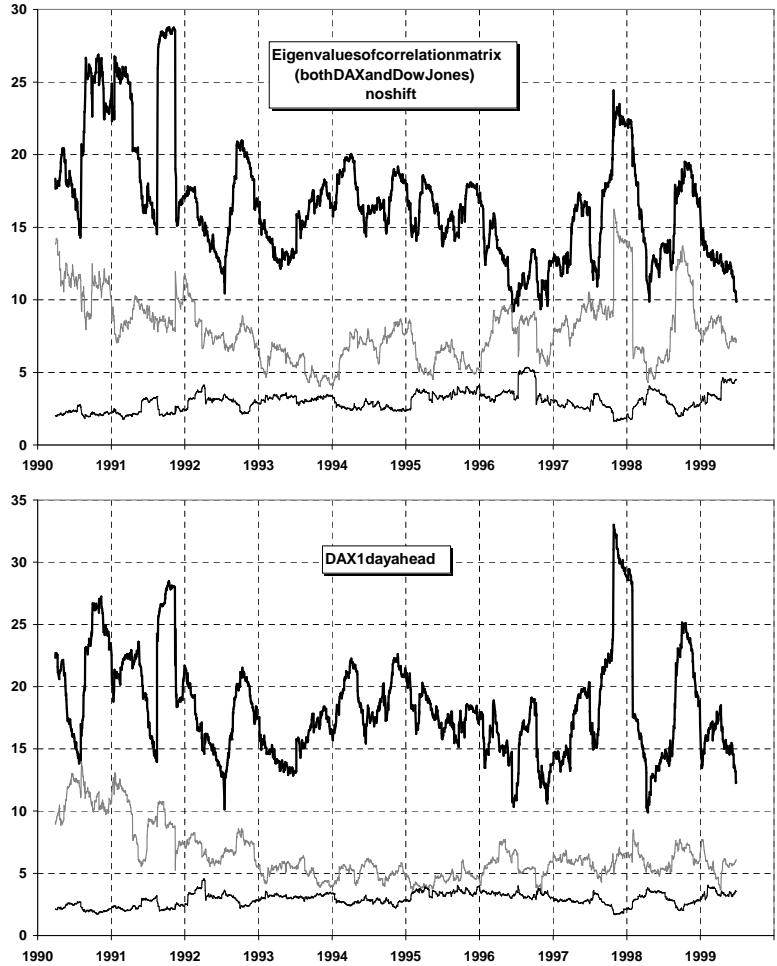}
\caption[]{Upper panel: Time-dependence of three largest eigenvalues
corresponding to the global (DAX + Dow Jones) correlation matrix ${\bf
C}_{\cal G}$ calculated from the time-series of daily price changes in the
interval of $T=60$ past trading days. Lower panel: Same as above but here
the DAX returns are taken one day ahead relative to the Dow Jones returns
when ${\bf C}^{\cal G}$ is constructed. }
\label{fig5}
\end{figure}

In our specific case of the two stock markets the corresponding global 
$({\cal G})$ correlation matrix ${\bf C}^{\cal G}$ can be considered
to have the following block structure:
\begin{eqnarray}
{\bf C}^{\cal G} =  
\left ( \matrix{ {\bf C}^{DAX,DAX} & {\bf C}^{DAX,DJ} \cr
                 {\bf C}^{DJ,DAX}  & {\bf C}^{DJ,DJ} } \right)
\end{eqnarray}

The time-dependence of the resulting three largest eigenvalues of ${\bf
C}^{\cal G}$ is illustrated in the upper panel of Fig.~\ref{fig5}.
In contrast to a single stock market case where the dynamics is typically
dominated by one outlying eigenvalue here one can systematically
identify two large eigenvalues. Both of them are always above the
range of variation of the remaining eigenvalues which stay confined to
the limits (eq.~\ref{eq:lmn}) prescribed by entirely random correlations
$(\lambda^{max}=4)$.

In fact the two largest eigenvalues represent the two stock markets
as if they were largely independent.
The time-dependences  
of the largest eigenvalue $\lambda^{\cal G}_1$ of ${\bf C}^{\cal G}$
and of the largest eigenvalue $\lambda^{DAX}_1$ of ${\bf C}^{DAX,DAX}$   
approximately coincide. The same applies to
the second largest eigenvalue $\lambda^{\cal G}_2$ of ${\bf C}^{\cal G}$
when compared to the largest eigenvalue $\lambda^{DJ}_1$ of ${\bf
C}^{DJ,DJ}$.
No explicit documentation is necessary, since this already can be seen
by comparing the present eigenvalues to those of Figs.~2 (middle panel)
and 4 (lower panel).

The above thus indicates that the two sectors represented by ${\bf
C}^{DAX,DAX}$ and by ${\bf C}^{DJ,DJ}$ respectively, remain practically
disconnected. Such a conclusion, however, is somewhat embarrassing,
because at the same time $\lambda^{\cal G}_1$ and $\lambda^{\cal G}_2$
(similarly as $\lambda^{DAX}_1$ and $\lambda^{DJ}_1$) go in parallel as
far as their time-dependence is concerned, especialy over the last few
years. Also the DAX and the Dow Jones increases and decreases,
respectively, display significant correlations. Both these facts point to
some sizeable correlations between the two markets. These become evident
when the correlation matrix ${\bf C}^{\cal G}$ is calculated from
$x_j^{DJ}(t)$ and $x_i^{DAX}(t+1)$, i.e., the DAX returns are taken one
day advanced relative to the Dow Jones returns. The resulting
time-dependence of the three largest eigenvalues is shown in the lower
panel of Fig.~\ref{fig5}. Now, except for the early 90's, one large
eigenvalue dominates the dynamics. This means that a sort of a single
common market emerges. Moreover, this common market also obeys the
characteristics observed before for the single markets. The collectivity
of the dynamics is weaker (smaller $\lambda_1^{\cal G}$) during increases
than during decreases.

\section{Summary}

In summary, the correlation matrix analysis of the stock market evolution
allows to quantify the co-existence of collectivity and noise and shows
that both are present. The majority of eigenvalues falls into limits
prescribed by random matrix theory. The largest one, however, represents a
collective state, whose time dependence provides arguments for a distinct
nature of the mechanism governing financial increases and decreases,
respectively. The increases are less collective by involving more
competition as compared to decreases. Such characteristics can even be
identified for a newly emerging global market in which the Dow Jones seems
to be leading.

\section{Acknowledgement}

We are grateful to the organizers of the Symposium for their invitation.
We thank the Symposium sponsors for financial support.

\clearpage
\addcontentsline{toc}{section}{Index}
\flushbottom
\printindex

\end{document}